\documentclass{ptephy_v1}
\usepackage{hyperref}
\hypersetup{
    colorlinks=true,
    citecolor=blue,
    linkcolor=blue,
    filecolor=black,      
    urlcolor=blue,
    pdftitle={RESTORE_ANN},
    pdfpagemode=FullScreen,
    }
\usepackage{color} 
\usepackage{kotex} 
\usepackage{ulem}
\usepackage{lineno}

\begin{document}

\title{Restoring the saturation response of a PMT using pulse-shape and artificial-neural-networks}

\author[1]{Hyun-Gi Lee}
\author[2,*]{Jungsic Park}
\affil[1]{Center for Precision Neutrino Research, Department of Physics, Chonnam National University,  Gwangju\,61186, Korea}
\affil[2]{Department of Physics, Kyungpook National University, Daegu\,41566, Korea\email{jungsicpark@knu.ac.kr}}

\begin{abstract}
The linear response of a photomultiplier tube\,(PMT) is a required property for photon counting and reconstruction of the neutrino energy. The linearity valid region and the saturation response of PMT were investigated using a linear-alkyl-benzene\,(LAB)-based liquid scintillator. A correlation was observed between the two different saturation responses, with pulse-shape distortion and pulse-area decrease. The observed pulse-shape provides useful information for the estimation of the linearity region relative to the pulse-area. This correlation-based diagnosis allows an \textit{in-situ} estimation of the linearity range, which was previously challenging. The measured correlation between the two saturation responses was employed to train an artificial-neural-network\,(ANN) to predict the decrease in pulse-area from the observed pulse-shape. The ANN-predicted pulse-area decrease enables the prediction of the ideal number of photoelectrons irrelevant to the saturation behavior. This pulse-shape-based machine learning technique  offers a novel method for restoring the saturation response of PMTs.

\end{abstract}

\subjectindex{C44, C50, H20}

\maketitle

\section{Introduction \label{sec1}}

A photomultiplier tube\,(PMT) amplifies the number of photoelectrons\,(NPE) released from a photocathode at each dynode and discharges them as a pulse of current. The linearity of the gain between the photocathode and the amount of output signal is a required property for counting NPE from the observed currents. It is known that saturation behavior appears when detecting an event of high-light intensities\,\cite{hamamatsu2007photomultiplier, wright2017photomultiplier} and the validity of the linearity depends on the environment\,\cite{babicz2019linearity,belver2018cryogenic}. Hamamatsu's 10-inch photocathode PMT\,R7081 is a widely used photon sensor specialized in neutrino experiments\,\cite{hama2017PHOTOMULTIPLIER, de2022double, ajimura2021jsns2, ANTARES:2005hwh, ahn2012observation}. Their linearity has been tested up to $\rm \sim 300\,(\rm 600)\,PE$ at a gain of $1\times 10^7\,(5\times 10^6)$\,\cite{bauer2011qualification,ajimura2021jsns2} and the characteristics of the saturation behavior have been investigated\,\cite{abbasi2010calibration}.

Reconstruction of neutrino events is typically performed from the NPE observed by each PMTs installed in the detector\,\cite{an2017measurement,galbiati2006time,qian2021vertex}. When it comes to reconstructing relatively high-energy particles, there are concerns regarding saturation behavior as the number of photoelectrons observed by a PMT\,(observed NPE) increases.  The saturation behavior is a major limitation in the reconstruction of neutrino produced by kaon-decay-at-rest\,$\rm (236\,MeV)$ in the ${\rm JSNS}^2$ experiment\,\cite{ajimura2021jsns2, ajimura2017technical} or passing through muon\,$\rm(\sim GeV)$ in the Double Chooz experiment\,\cite{ardellier2006double}. For each case, a  diagnosis of the linearity range or an understanding of the saturation response is required for the reconstruction of the events occurring in these energy regions.

The observed NPE associated with the reconstructed particle energy also depends on the detector size and light-collection method. In the very-short baseline reactor $\bar{\nu}_e\,\rm(\sim few\,MeV)$ experiments, several PMTs were installed in each segmented cell with different baselines, and the observed NPE reached $\rm \sim 500\,PE/MeV$\,\cite{ashenfelter2018performance}. These types of experiments prove the existence of sterile neutrinos\,\,based on the comparison of the observed $\bar{\nu}_e$ energy spectra for the segmented region with different baselines\,\cite{andriamirado2021improved, almazan2020improved, serebrov2021search}. Global analyses of available $\bar{\nu}_e$ disappearance data with the existence of sterile neutrino have indicated that the best-fitted value of $(\sin^{2}{\theta}_{14},\,\Delta m^{2}_{41})$ is $(0.009,1.3)$\,\cite{dentler2018updated}. An identical response for each detector is required to prove the energy-dependent disappearance of the $\bar{\nu}_e$ with such a small mixing angle\,\cite{seo2018spectral,an2017measurement}. The saturation behavior of PMTs may spoil the identical energy scale response for each cell, which gives ambiguity in the precision comparison of the prompt energy spectra or reduce sensitivity in the sterile neutrino search\,\cite{andriamirado2021improved, almazan2020improved}.

In this study, we investigate a possible diagnosis and restoring saturation response of R7081 up to $\rm \sim4000\,PE$ using artificial-neural-networks\,(ANNs) and the observed pulse-shape. A correlation between the two different types of saturation responses, pulse-shape distortion and pulse-area decrease, was observed. The obtained correlation was employed to train the ANN to recover the reduced pulse-area from the observed pulse-shape, which contains \textit{in-situ} information on the pulse-area decrease. Section\,\ref{sec2} describes the experimental setup for the saturation response measurements. Section\,\ref{sec3} elucidates the measurement of gain and saturation response. Section\,\ref{sec4} provides details on the structure and training of the ANN. Section\,\ref{sec5} presents the training results and restoration of the saturation response from the saturated pulse-shape. Finally, Section\,\ref{sec6} summarizes the study and highlights the major conclusions drawn from the obtained results.

\begin{figure}[!h]
\includegraphics[width=15.cm]{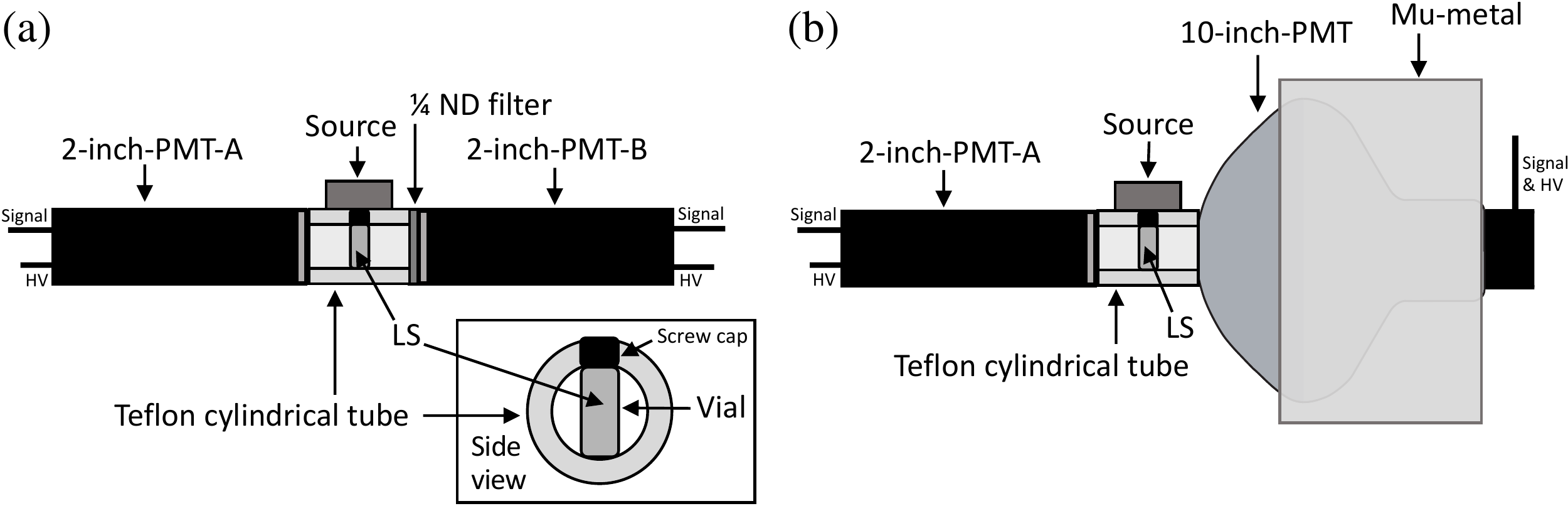}
\centering
\caption{Experimental setup for the PMT response measurement. (left):\,Test setup used for the linearity test of the 2-inch-PMT-A is drawn in the left. The 2-inch-PMT-B receives scintillation light attenuated by a $\rm 1/4$\,\,neutral-density-filter and monitors the linearity of the 2-inch-PMT-A. (right):\,Experimental setup for the saturation response measurement of the 10-inch-PMT. The 2-inch-PMT-A\,(10-inch-PMT) observed the scintillation events, without\,(including) the saturation.}
\label{fig_1_setup}
\end{figure}

\section{Experimental setup \label{sec2}}

The saturation response of R7081 was investigated using the setups shown in Figure\,\ref{fig_1_setup}. Two different types of PMTs, 10-inch-PMT\,(Hamamatsu R7081) and 2-inch-PMT\,(Hamamatsu H7195) were used to detect the scintillation events. The 2-inch-PMT was designed to detect the scintillation events without the saturation,  whereas the 10-inch-PMT coincidently detected the same scintillation events, including the saturation response.

The linearity\,(saturation response) of the 2-inch-PMT\,(10-inch-PMT) was measured using the configuration shown in Figure\,\ref{fig_1_setup}\,left\,(right). A mu-metal was used to shields the 10-inch-PMT from the geomagnetic field to improve its collection efficiency\,\cite{devore2014light}. A glass vial contained the liquid-scintillator\,(LS), which yielded the scintillation light from ionization. The $\gamma$-rays emitted from the radioactive sources of $^{137}{\rm Cs}\,(0.66 \rm\, MeV)$ and $^{35}{\rm Cl}\,(\rm \sim8\, MeV)$ transferred their energy to the electrons of the LS via Compton scattering and further ionized the scintillator. The emitted scintillation photons were reflected in the Teflon cylindrical tube, and 10-inch or 2-inch-PMT coincidently detected these photons with different quantum and collection efficiencies. The Teflon cylindrical tube was made of Polytetrafluoroethylene\,(PTFE) and had a length of 7.5\,cm with an inner diameter of 2\,inches. The gain of each PMT was adjusted to the typical gain provided by the manufacturer$\,(\sim3\times10^6\,\,\rm\, for\,\,H7195,\,\sim1\times\,10^7\,\,for\,\,R7081)$\,\cite{hama2017PHOTOMULTIPLIER}, and the observed pulse-area was divided into gain to determine the observed NPE.

\begin{figure}[!h]
\includegraphics[width=11cm]{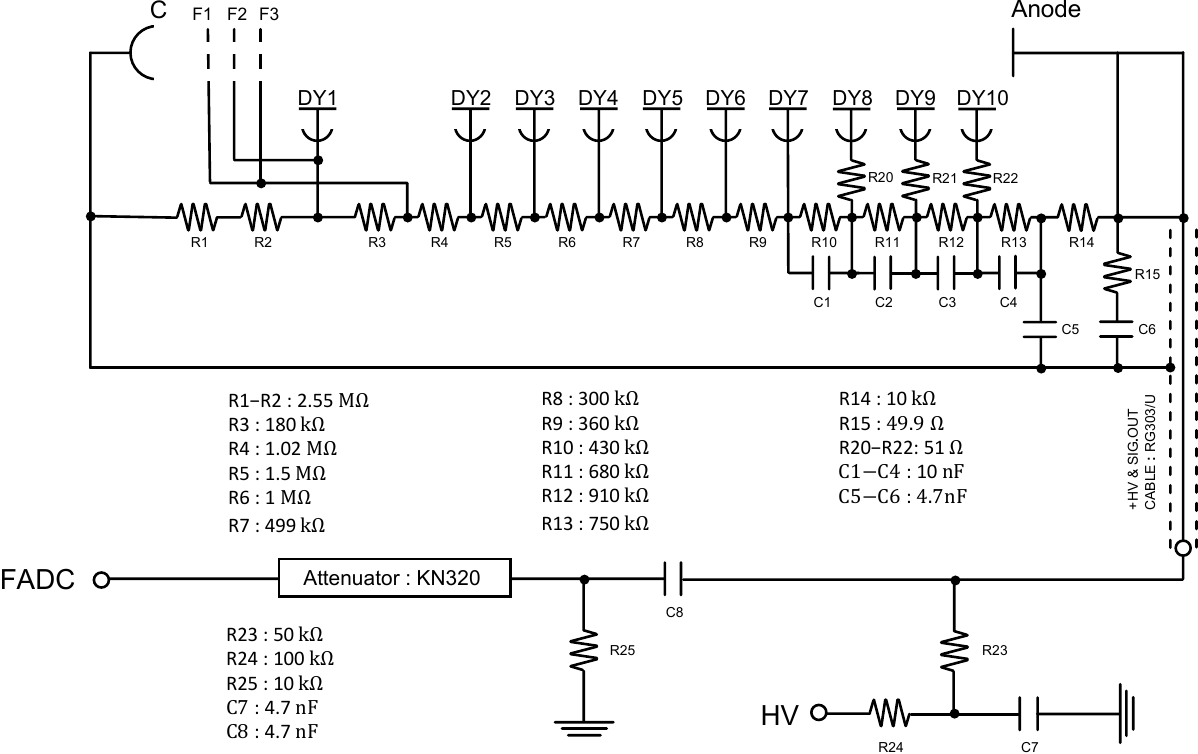}
\centering
\caption{Circuit diagram of R7081. The total resistance between the dynodes is 12.7$\rm\,M\Omega$, and the distribution ratio of the dynode resistance is (16.8-0.6-3.4-5-3.33-1.67-1-1.2-1.5-2.2-3-2.4)\,\cite{Ma:2009aw}.}
\label{fig_2_circuit}
\end{figure}

\vspace{-0.4cm}

 An LS is a mixture of an organic solvent, a fluor, and a wavelength shifter. Linear-alkyl-benzene\,$(\rm LAB, C_{n}H_{2n+1}$-$\rm C_{6}H_{5}, n = 10$-$13)$ is a widely used as a solvent for LS owing to its relatively high light yield and environmentally friendly characteristics\,\cite{park2013production, park2019production, beriguete2014production, abusleme2021optimization, buck2019production, kim2016development}. In this study, 2,5-diphenyloxazole\,$(\rm PPO, C_{15}H_{11}NO)$ and 1,4-bis(2-methylstyryl)benzene\,$(\rm bis$-$\rm MSB, C_{24}H_{22})$ were adopted as the fluor and the wavelength shifter, respectively. The LS was synthesized by dissolving a $3\,\rm g/L$ of PPO  and $30\,\rm mg/L$ of bis-MSB to LAB, which is an aromatic solvent. The primary decay time of the LS is about $\sim$4\,$\rm ns$\,\cite{Zhong2008decaytime}.

\vspace{0.cm}

As described in Section\,\ref{sec1},  the NPE released from the photocathode is multiplied at each dynode and discharged as  current pulses. Figure\,\ref{fig_2_circuit} presents a relevant circuit diagram for the 10-inch-PMT, which contained an attenuator\,(Kaizuworks KN320), a flash analog to digital converter\,(FADC), and a high-voltage power supply\,(ORTEC 556). The discharged pulses were attenuated to adjust the dynamic range of the electronics and digitized by Notice FADC400, which is a 10-bit, 400\,MHz/s, $\rm \pm1V_{pp}$ FADC\,\cite{FADC400}.

\section{Measurements \label{sec3}}

\begin{figure}[h]
\includegraphics[width=5.cm]{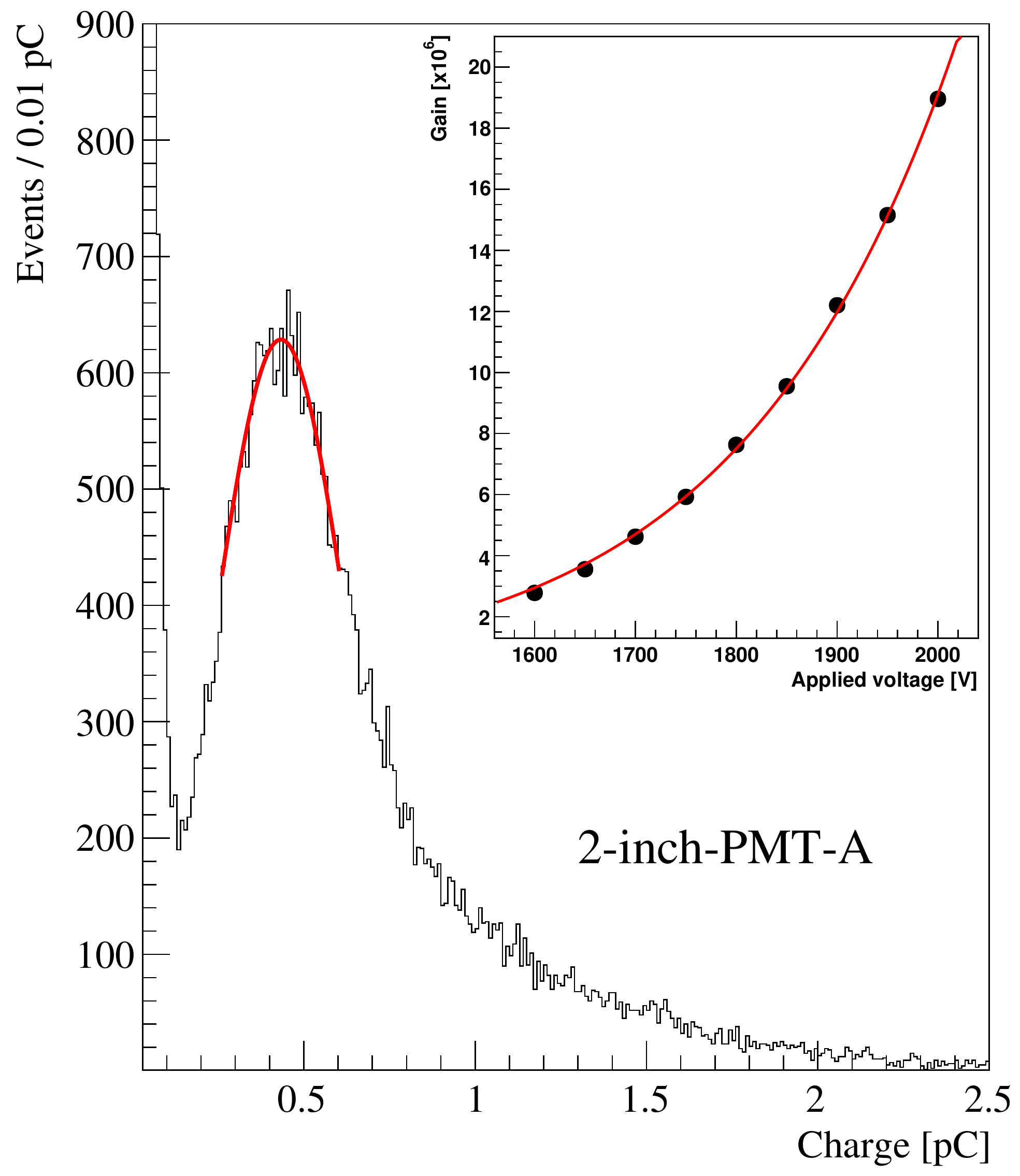}
\includegraphics[width=5.cm]{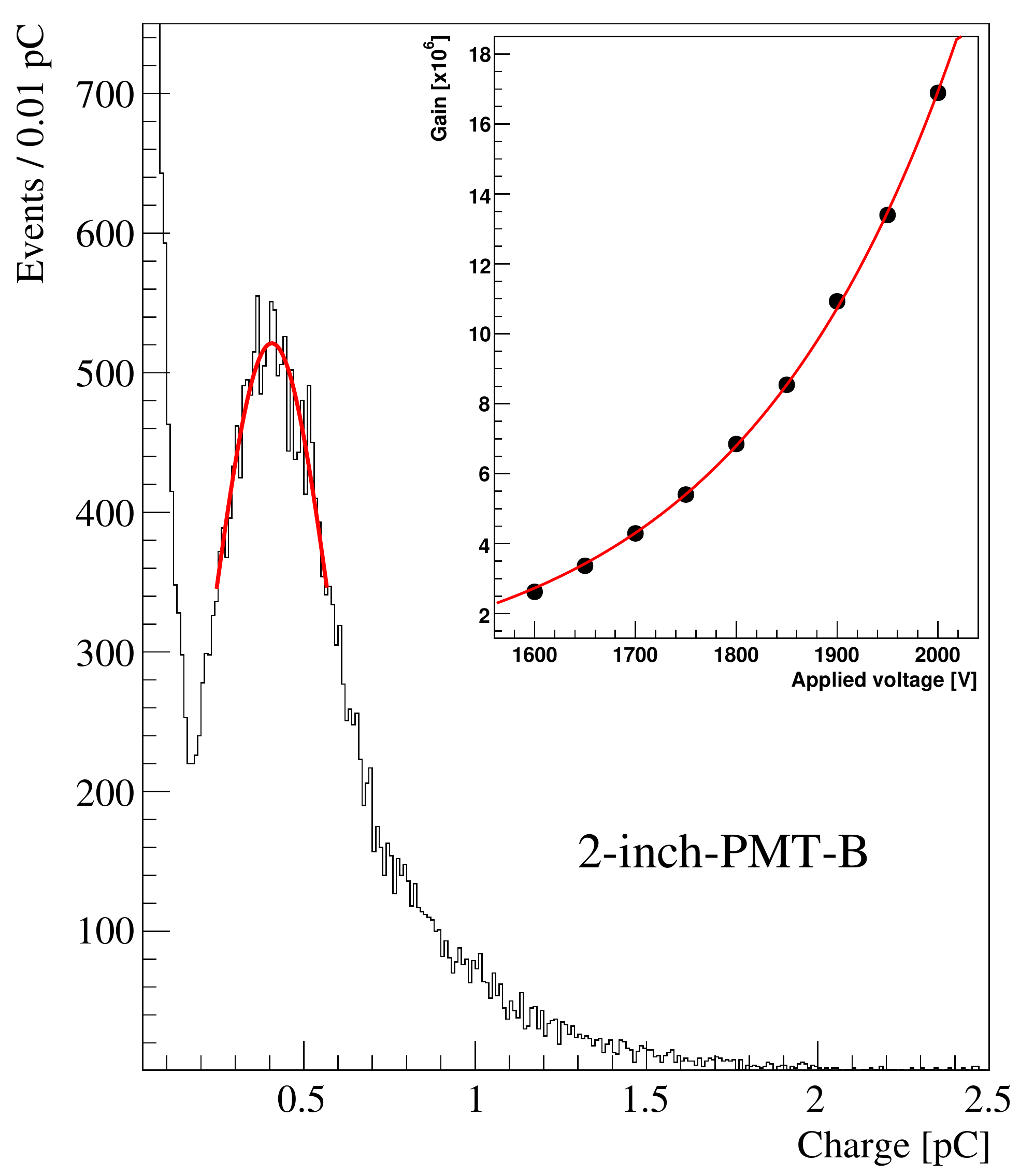}
\includegraphics[width=5.cm]{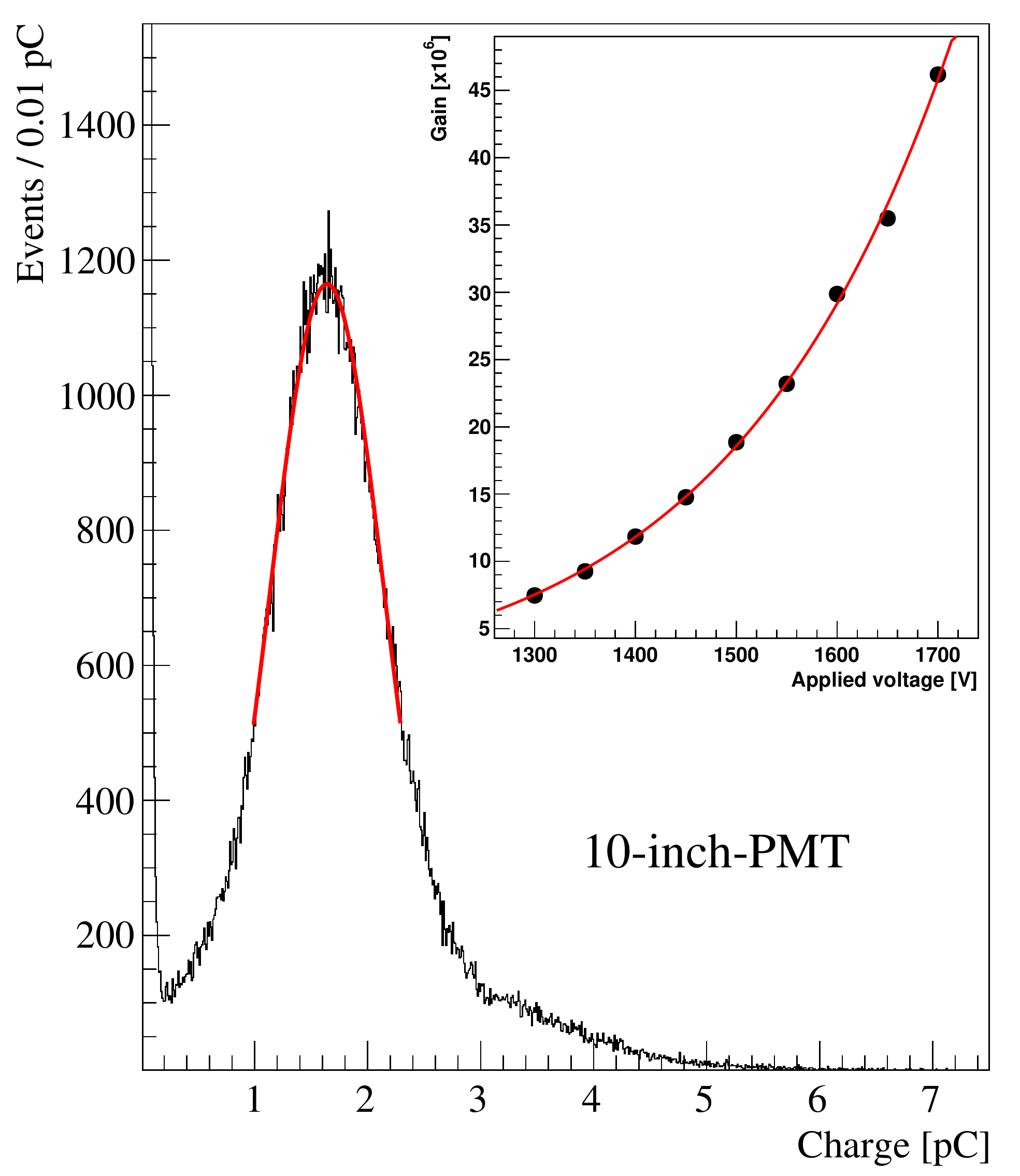}
\centering
\caption{Measurement of the gain of each PMTs. The charge distribution of the 2-inch-A, 2-inch-B and 10-inch-PMT are presented on the left, middle, and right, respectively. The insets display the exponential increase in the observed gain with the increasing applied voltage.} 
\label{fig_3_gain}
\end{figure}

The gain of each PMT was measured using an attenuated laser light source with an external trigger.  For obtaining  the single photoelectron\,(SPE) charge distribution, a 440\,nm laser light\,(OPG-NIM-440) was attenuated by  neutral-density-filter\,$(0.01\%)$. Figure\,\ref{fig_3_gain} shows the observed charge distribution from the PMT with the light source. The inset indicates the exponential increase in the SPE as a function of the applied voltage. The adjusted gains are $2.99 \times 10^6$\,(2-inch-A), $2.78 \times 10^6$\,(2-inch-B), and $1.02 \times 10^7$\,(10-inch) with the applied voltages of 1600, 1600, and 1365\,V, respectively.

\vspace{-0.1cm}

\begin{figure}[]
\centering
\includegraphics[width=14.8cm]{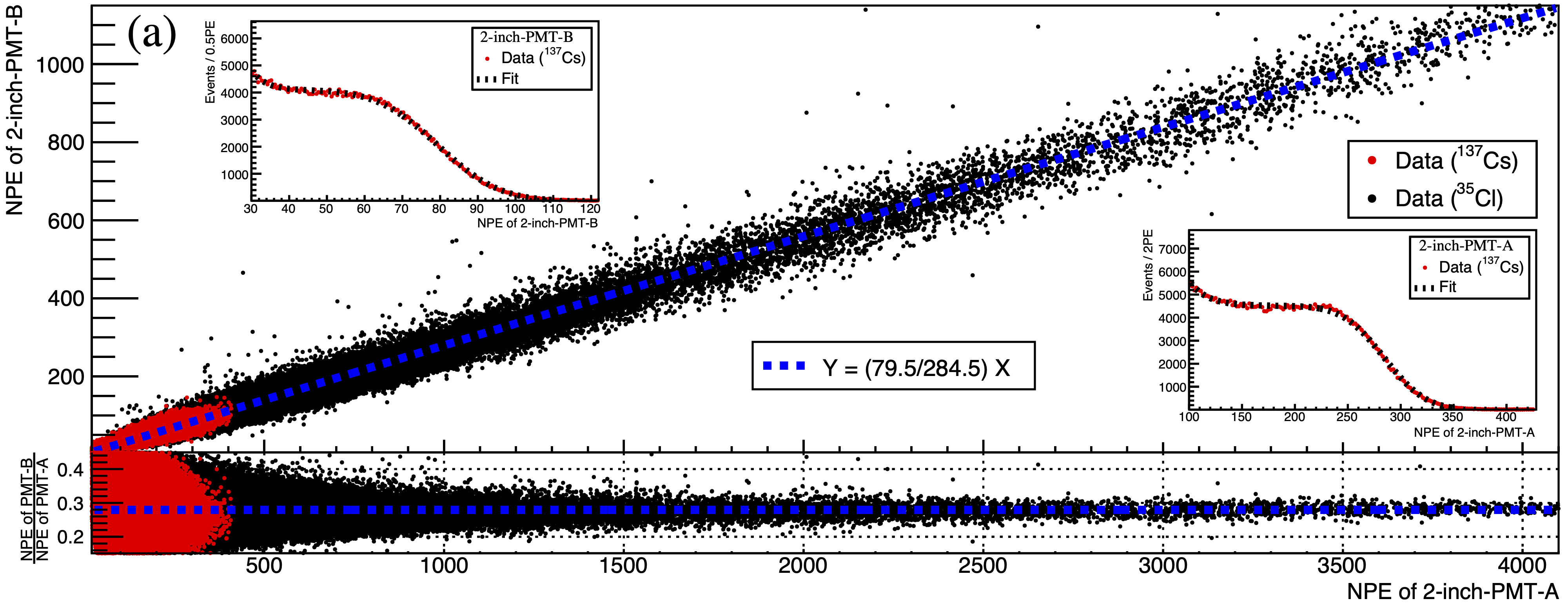}\\
\centering
\includegraphics[width=14.8cm]{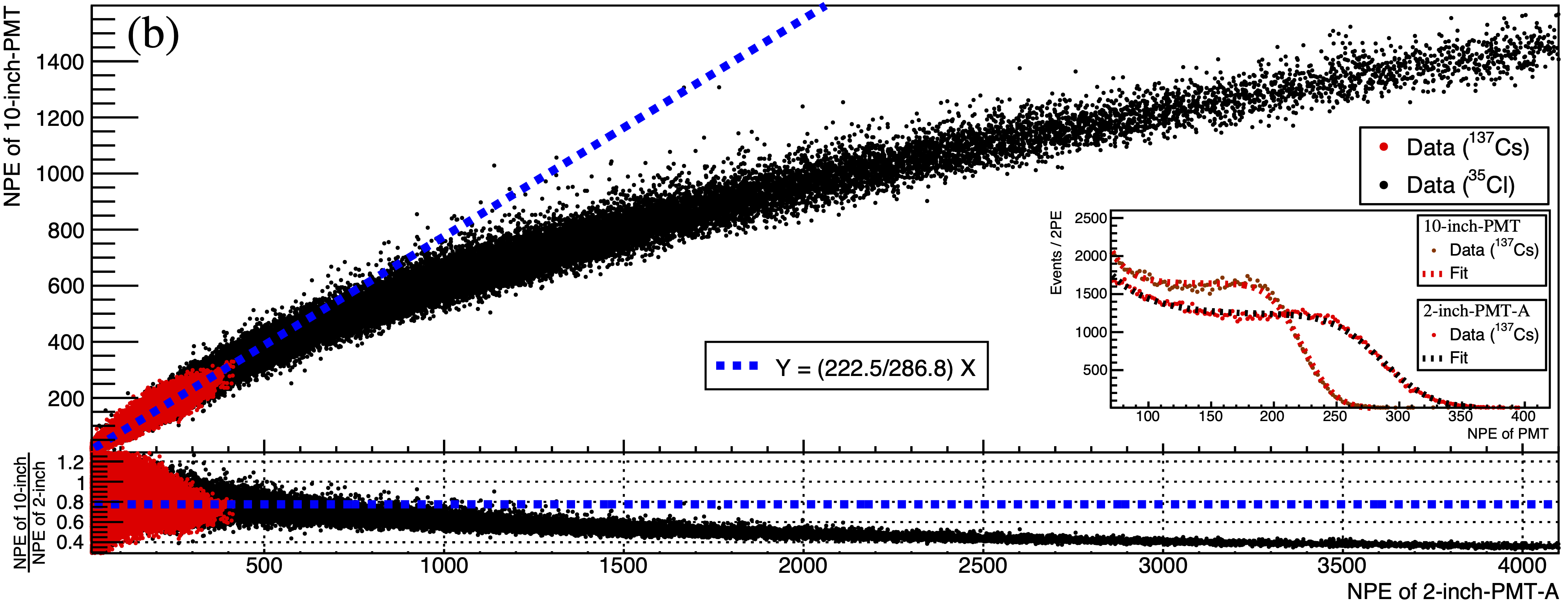}\\
\centering
\caption{Obtained PMT response at the typical gain using the setup shown in Figure\,\ref{fig_1_setup}. Response of  the 2-inch-PMT-A and 2-inch-PMT-B\,(10-inch-PMT) drawn in the Figure\,\ref{fig_4_PE_response}a\,(b). The blue dashed-line indicates the linear-model and their slope is determined from the ratio of the observed Compton edges\,$\rm (79.5/284.5\,(222.5/286.8)\,[NPE/NPE])$.\label{fig_4_PE_response}} 
\end{figure}

As described in the previous section, the scintillation events are detected by the coincidence of two PMT with different efficiencies. The relative difference between the photon detection efficiencies of the two PMT shown in Figure\,\ref{fig_1_setup} was obtained by the comparing observed Compton edges\,\cite{hristova1990compton}. In Figure\,\ref{fig_4_PE_response}, the red data points illustrates the NPE observed by  the two PMTs obtained using a $^{137}{\rm Cs}\,(0.66 \rm\, MeV)$ $\gamma$-ray source. A one-dimensional projection of the scatter plot is presented in the inset, and their distribution is fitted with an empirical formula of Error\,+\,Exponential function\,$(y= {\rm P_0} \cdot  \left[ 1 - \rm erf \left( (x - {\rm P_1})/{\rm P_2} \right)  \right]  +  {\rm P_3} \cdot \exp \left( -{\rm P_4} \cdot x \right))$ to find the edge of ${\rm P_1}$\,\cite{xu2017new,antcheva2011root}.  The blue-dashed line in Figure\,\ref{fig_4_PE_response} indicates the linear-model\,$\rm (Y\propto X)$, which has a slope of the ratio of the observed edge between the two PMT. From the consistency between the data and model in the upper figure, the ratio of the observed edge between the two PMTs is employed as an efficiency correction coefficient. By multiplying the ratio of the obtained edges with the observed NPE of 2-inch-PMT-A, the ideal NPE of the 10-inch-PMT irrelevant to the saturation behavior is estimated.

The responses of the PMTs at a higher scintillation light intensity were obtained using a $^{35}{\rm Cl}\,(\rm \sim8\, MeV)$ $\gamma$-ray source\,\cite{mao2020high}. In Figure\,\ref{fig_4_PE_response}, the black data points shows a comparison between the observed NPE of the two PMTs at a larger NPE. Figure\,\ref{fig_4_PE_response}a compares the NPE observed by the 2-inch-PMT-A and  2-inch-PMT-B from the same scintillation events. To test the linearity of the 2-inch-PMT-A, a $\rm 1/4$\,\,neutral-density-filter was installed on the 2-inch-PMT-B. No saturation behavior is evident for $\rm \sim 4000\,PE$ observed by the 2-inch-PMT-A. Figure\,\ref{fig_4_PE_response}b compares the response of the 2-inch and 10-inch-PMTs. As the NPE increases, the saturation behavior of the 10-inch-PMT becomes distinct.

\begin{figure}[]
\includegraphics[width=7.8cm]{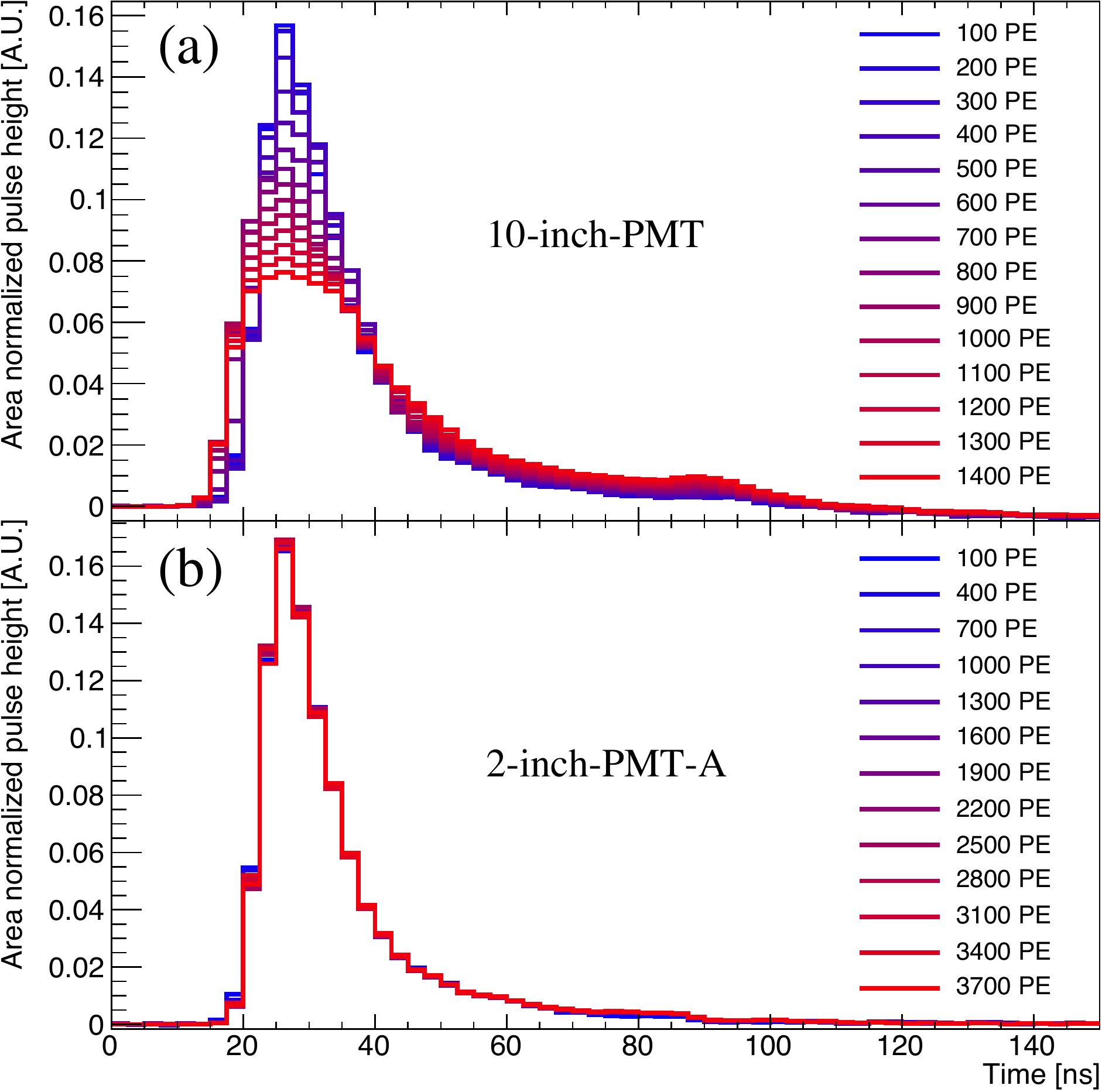}\,\,\,
\includegraphics[width=7.23cm]{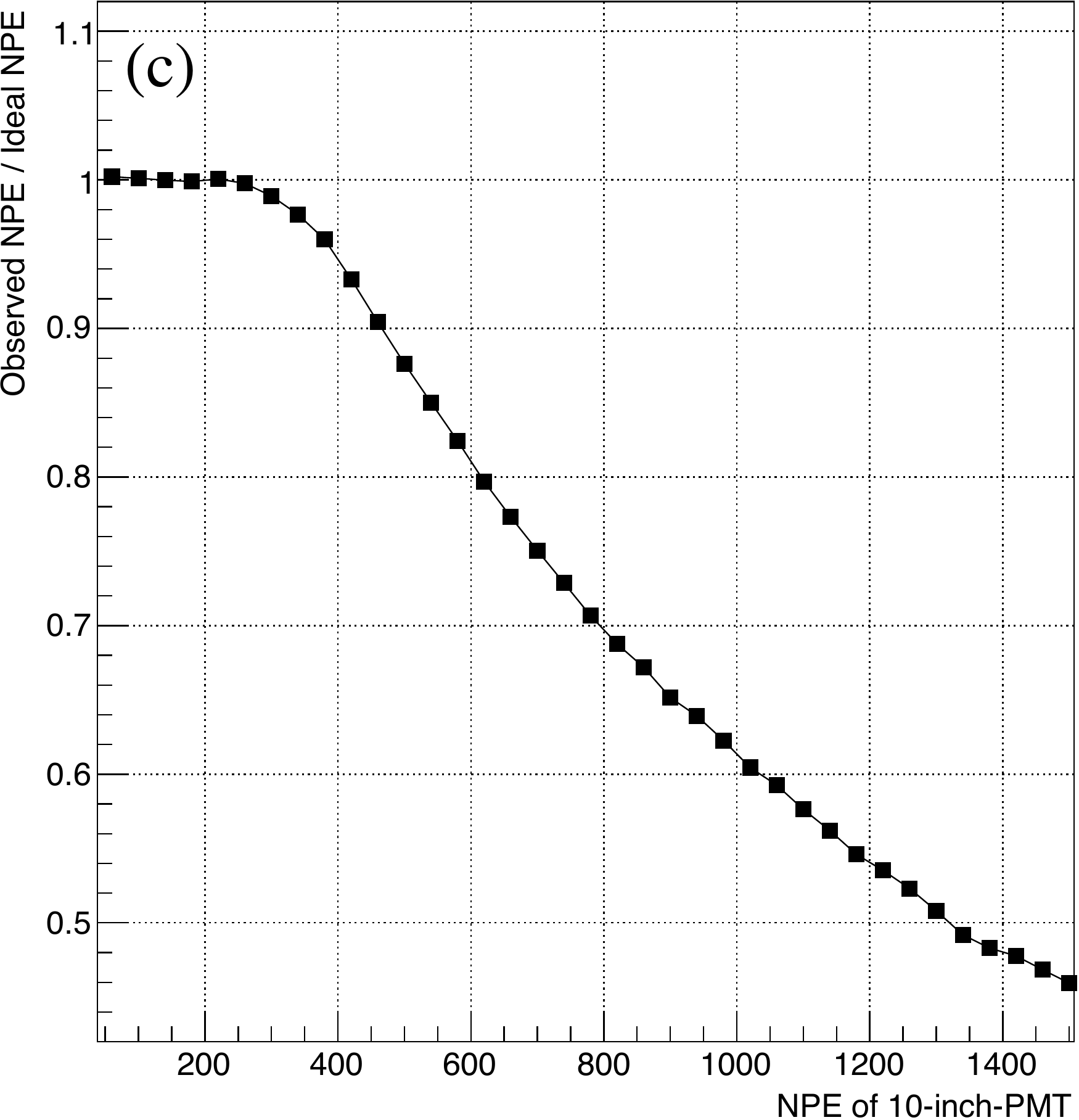}
 
\centering
\caption{Observed pulse-shape and pulse-area saturation response. (left):\,Observed pulse-shape distortion\,(linearity) of the 10-inch-PMT\,(2-inch-PMT-A). (right):\,Observed decrease in pulse-area shown by the 10-inch-PMT. The ideal NPE of the 10-inch-PMT which is irrelevant to the saturation behavior is obtained based on the NPE of 2-inch-PMT-A. The responses were measured with respect to the observed NPE for each PMT.}
\label{fig_5_correlation}
\end{figure}

In the operation of a PMT with a high-intensity light pulse, the electron trajectory between the dynodes is distorted by the space charge effect caused by the repulsive force of the electron cloud\,\cite{wright2017photomultiplier, hamamatsu2007photomultiplier}, and the saturation behavior of the PMT appears as a distortion of the pulse-shape and simultaneous decreases in the pulse-area\,\cite{abbasi2010calibration, babicz2019linearity}. Both the saturation responses were observed in the 10-inch-PMT and are presented in Figure\,\ref{fig_5_correlation}. The left side of Figure\,\ref{fig_5_correlation} shows the accumulated pulse-shape of the scintillation events with respect to the observed NPE of each PMT. The measured pulse-shape contains a rise, peak, and tail\,\cite{jeong2020pulse}. Figure\,\ref{fig_5_correlation}a displays the distortion of the pulse-shape of 10-inch-PMT appears as a decrease in the peak. The distortion starts around $\rm \sim 300\,PE$. On the other hand, in the Figure\,\ref{fig_5_correlation}b, a similar pulse-shape is observed for the 2-inch-PMT irrespective of the observed NPE. Figure\,\ref{fig_5_correlation}c presents the pulse-area decrease responses for the 10-inch-PMT. A deviation from the ideal response of more than 1\% appears at $\rm \sim 300\,PE$, and this deviation  increases as a function of the NPE of the 10-inch-PMT. The response is obtained by averaging the ratio between the ideal and observed NPE according to the NPE of the 10-inch-PMT.

An increase in the pulse-shape distortion and the pulse-area decrease was observed as a function of the observed NPE of 10-inch-PMT. Note that the pulse-shape distortion of the 10-inch-PMT is observed only using the 10-inch-PMT information, without comparing its response to that of the 2-inch-PMT. The correlation observed between the pulse-shape distortion and pulse-area decrease suggests the possibility of diagnosing the linearity range or predicting a decrease in the pulse-area from the observed pulse-shape.

\vspace{-0.cm}

\section{Training of artificial-neural-network\,(ANN)\label{sec4}} 

ANNs are widely used machine learning algorithms that can extract correlated features from an input of a higher dimension. A typical ANN consists of repeated layers of perceptron and a non-linear activation function. An output of the $i$-th perceptron in the $N$-th layer is feed forward to the $j$-th perceptron in the $(N+1)$-th layer with a connection strength of $w_{ij}$. Higher-dimension features are extracted as the layer is repeated, and the extracted features are feed forward to the next layer. The ANN has numerous free parameters for the unfixed $w_{ij}$, and multiple sets of data-label pairs are required to train these parameters. Further details on ANNs are provided in Ref.\,\cite{jain1996artificial}.

\begin{figure}[!h]
\includegraphics[width=7.2cm]{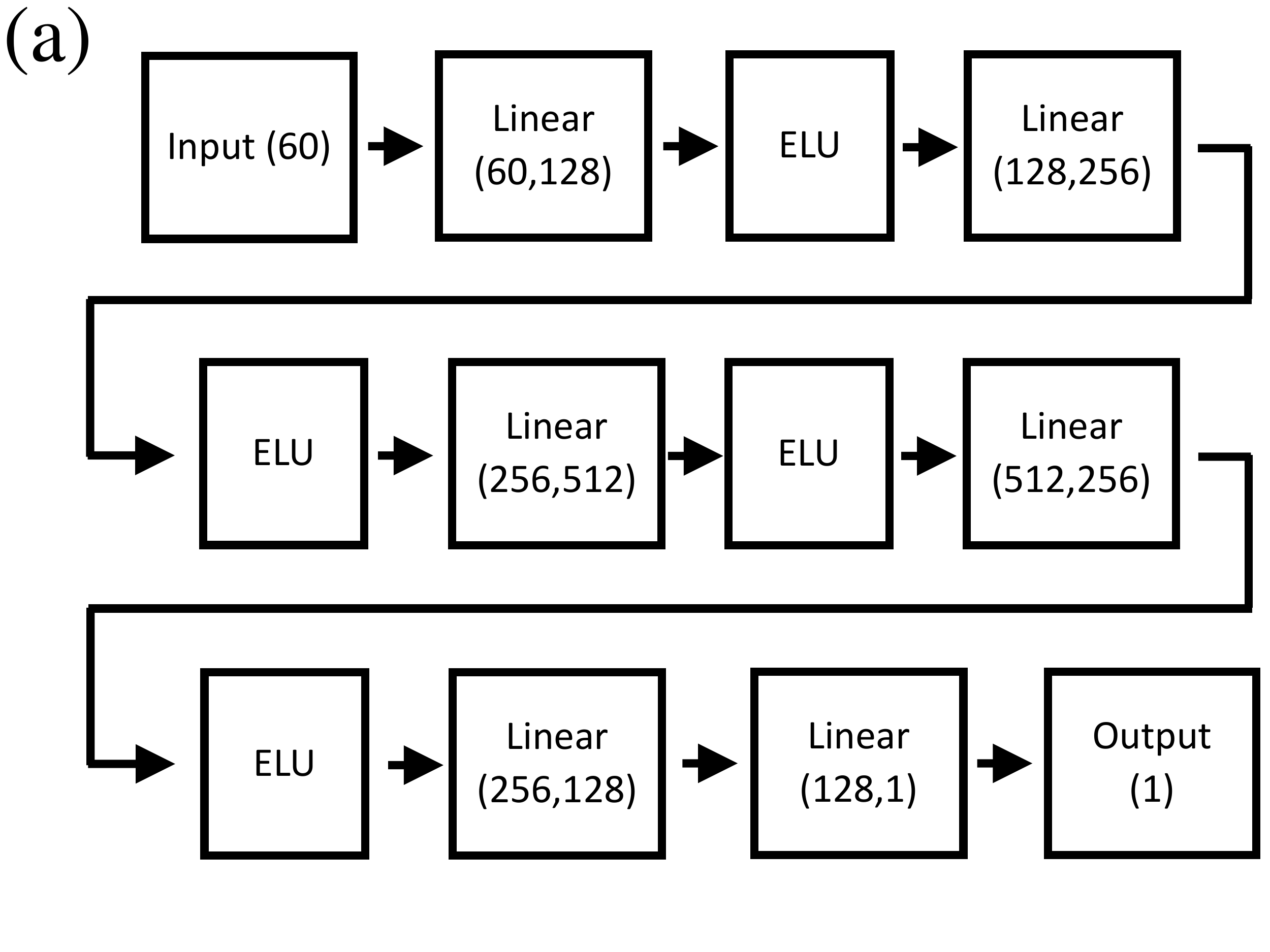} \,\,
\includegraphics[width=7.7cm]{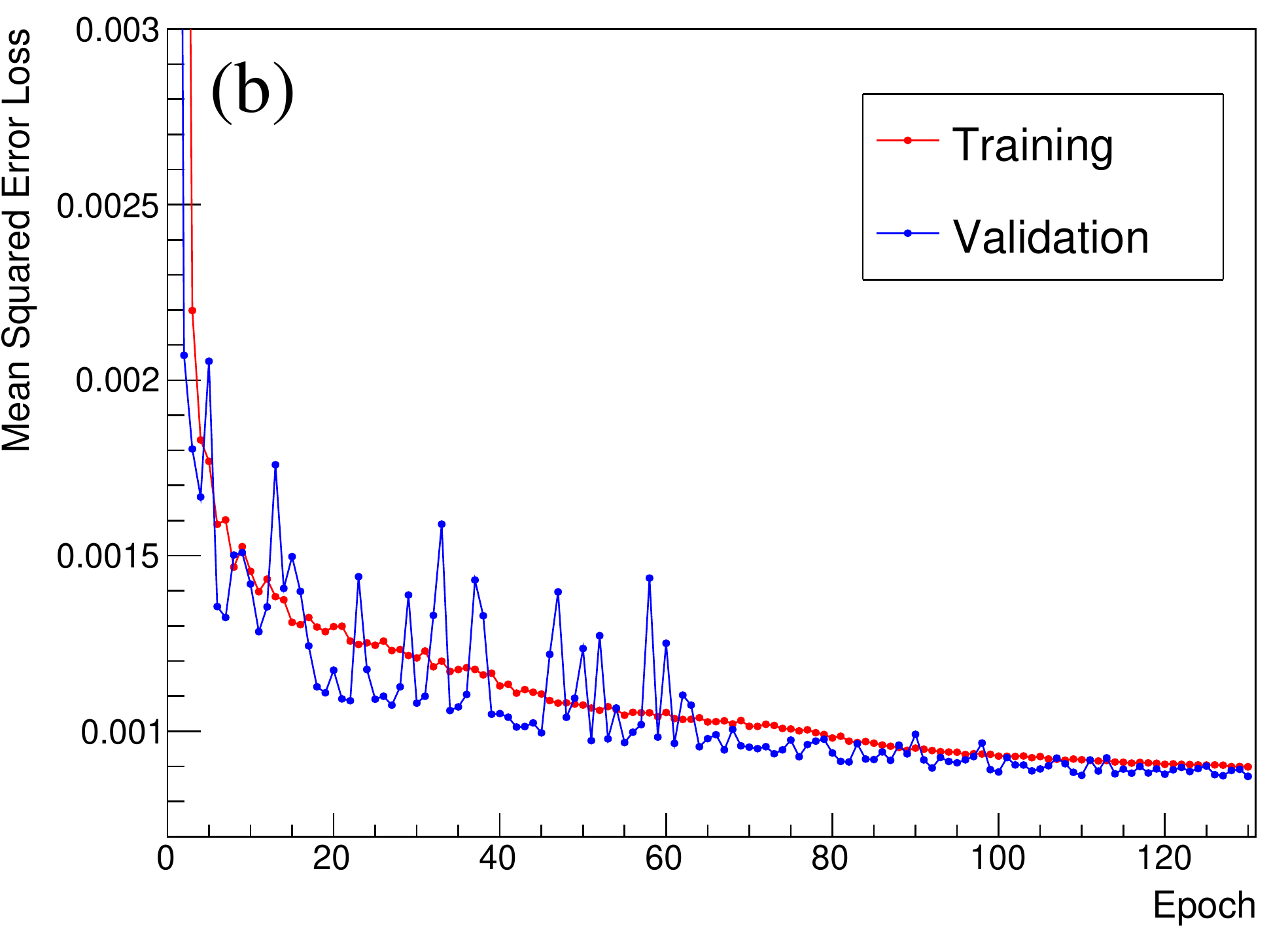}
\caption{Structure and training of the ANN. (left):\,Schematic diagram of the ANN used to predict the pulse-area saturation from the observed pulse-shape. (right):\,Decreasing in MSE  loss of the training and the validation sets during the epochs. The decreasing of the loss saturates at an epoch of 130.}
\label{fig_6_training}
\end{figure}

The ANN was trained using the observed correlation between the pulse-shape distortion and pulse-area decrease. Figure\,\ref{fig_6_training}a shows the structure of the ANN used for predicting the pulse-area decrease from the observed pulse-shape. The deep-learning framework PyTorch\,\cite{paszke2019pytorch} was adopted to construct the ANN model. The input is provided to 60\,nodes, and each node represents a single 2.5\,$\rm ns$ bin of the area-normalized pulse-shape. The area-normalized pulse-shape is then feed forward to the next layer by the exponential linear unit\,(ELU), which is an non-linear activation function\,\cite{clevert2015fast}, and repeated several times to extract the non-linear features corresponding to the pulse-area decrease. The extracted features are transferred to the linear layers to predict the decreased ratio of the observed pulse-area for restoration. 

\vspace{-0.2cm}

\vspace{-0.2cm}
 
As demonstrated in Figure\,\ref{fig_5_correlation}, the saturation response appears as  a distortion of  the pulse-shape and  a simultaneous decrease in the pulse-area. The free parameters in the ANN are trained by the number of input-label pairs obtained from the scintillation events. The input is the pulse-shape distortion response, which is given by the area-normalized pulse shape with 60 bins, while the label is inverse of the pulse-area decrease response defined by Ideal NPE/Observed NPE or $\rm Q/(Q- \Delta Q)$. Here, Ideal NPE is denoted as $\rm Q$, which is ideal charge from the pulse-area irrelevant to the saturation response. On the other hand, Observed NPE is denoted as $\rm Q-\Delta Q$, where $\rm \Delta Q$ is the reduced charge from the pulse area decreased by the saturation response.

To train, validate and test the ANN model, $2.0\times10^5$  input-label pairs were selected at NPE larger than 300 for the restoration. For each  event, the normalized pulse-shape of the 10-inch-PMT was used as the input. The corresponding label was determined from the observed pulse-area decrease response in Figure\,\ref{fig_5_correlation}c and the observed NPE of the 10-inch-PMT. The inverse of the pulse-area decreased ratio was adopted as a label\,$\rm (label = Ideal\;NPE/Observed\;NPE$ or $\rm Q/(Q- \Delta Q))$  and used as the pulse-area restoration coefficient for the decreased pulse-area. The splitting ratio of the training, validation and test datasets was 2:1:2. The correction should not be performed in the region where the absence of the saturation response and NPE smaller than 300 have been excluded for the restoration. 

Figure\,\ref{fig_6_training}b depicts the training of the ANN over each epoch. The Adam optimizer\,\cite{chilimbi2014project} and the Mean Squared Error\,(MSE) loss\,\cite{james1992estimation} were  utilized during the training process. The initial learning rate\,$(5\times 10^{-4})$ was gradually decreased at a rate of $\rm 0.975^{epoch}$ by using the learning rate scheduler.  During each epoch, the connection strengths represented as $w_{ij}$ between the perceptrons were optimized and leading to a decrease in the loss. The ANN was trained for a total of 130\,epochs.

\section{Results \label{sec5}}

\begin{figure}[!h]
\includegraphics[width=8.45cm]{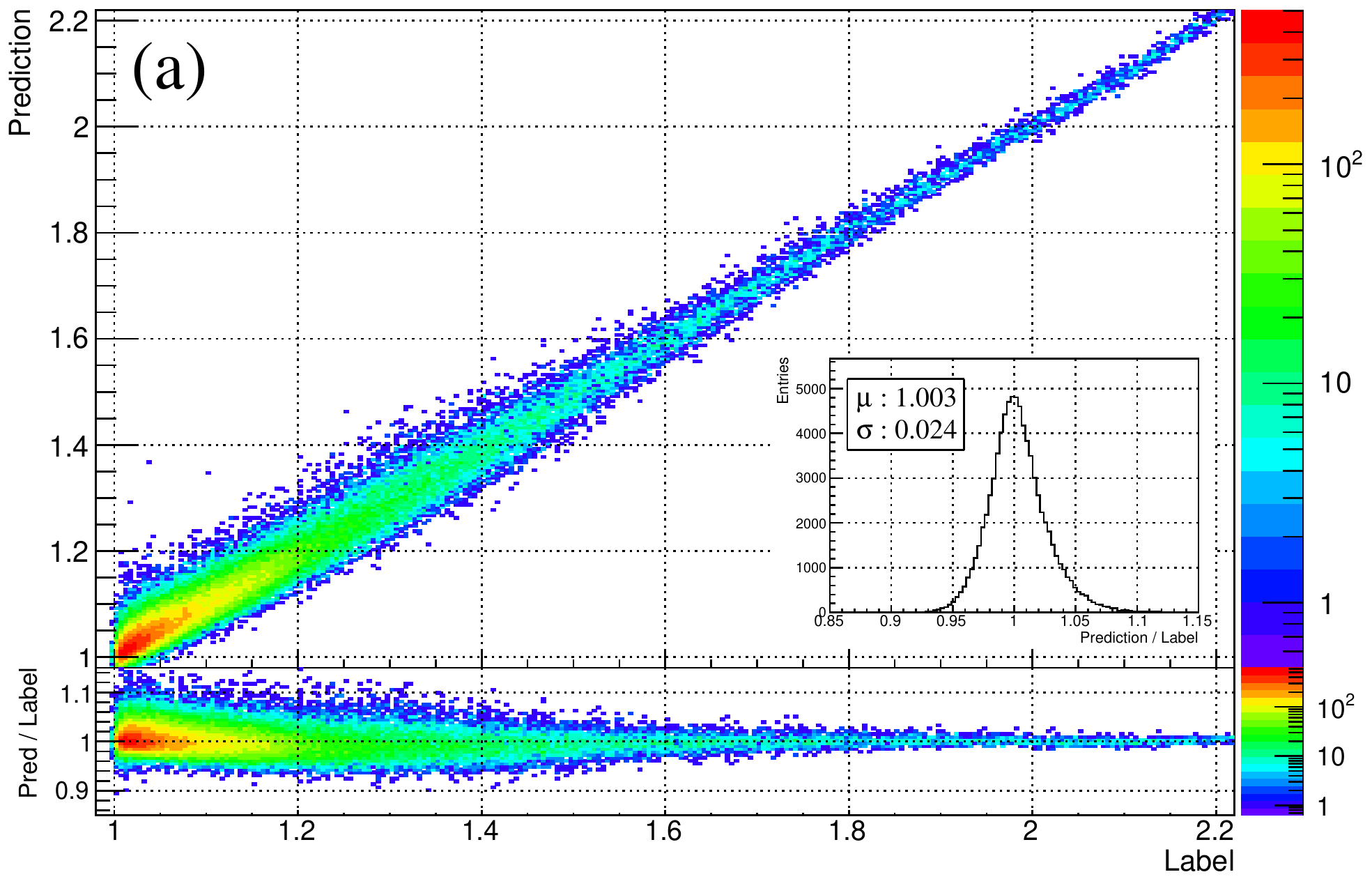}\,
\includegraphics[width=6.5cm]{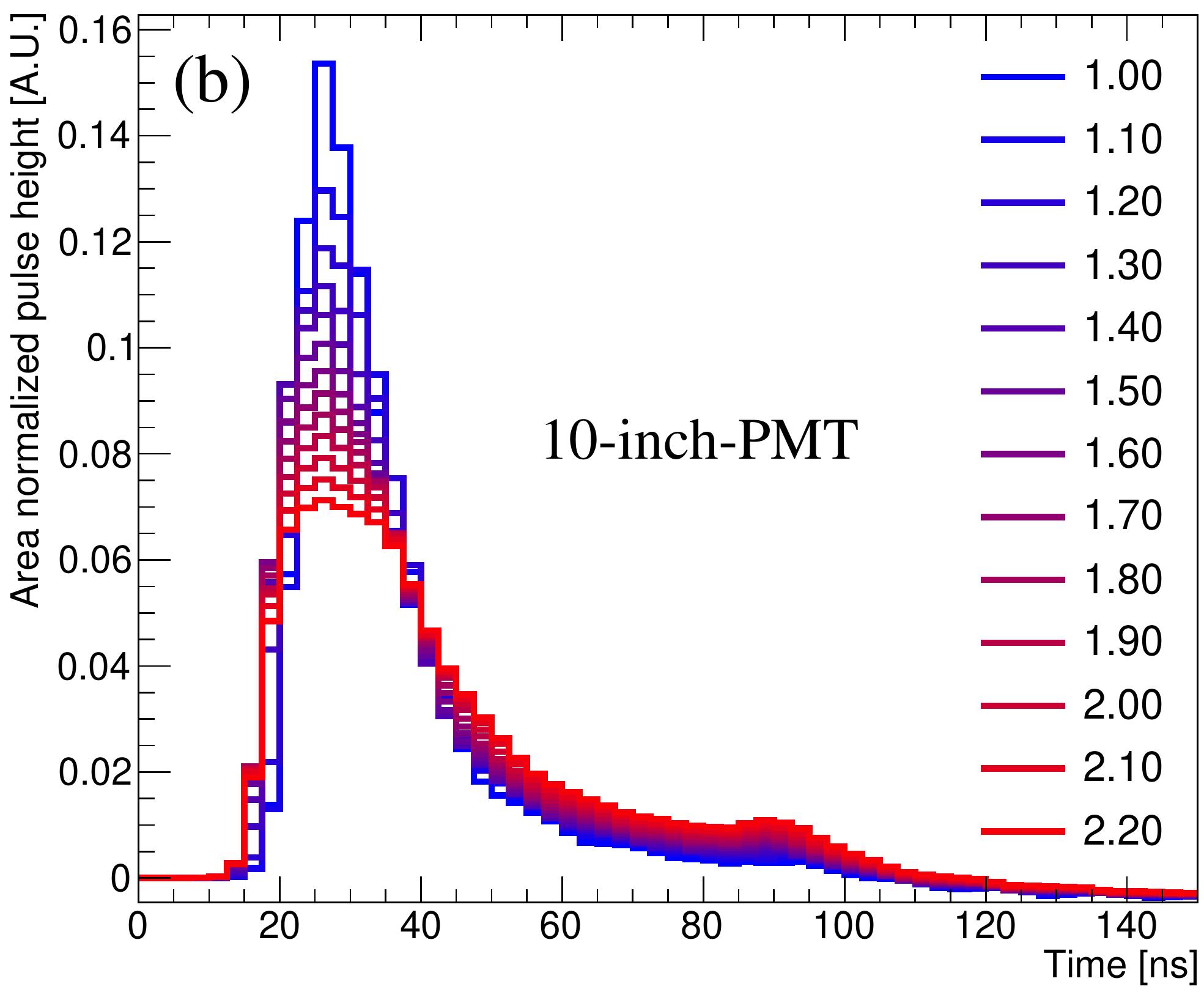}
\centering
\caption{Validation of the trained ANN using the test dataset. The trained ANN predicts the restoration coefficient for the decreased pulse-area from each observed pulse-shape. (left):\,Comparison of the labels and ANN predictions. (right):\,Classified  pulse-shape by the trained ANN for each prediction.}
\label{fig_7_training_result}
\end{figure}

The trained ANN was tested using the test dataset. The restoration coefficient corresponding to the decreased pulse-area was predicted by the trained ANN from the observed pulse-shape. A clear correlation between the ANN prediction and the label is shown in Figure\,\ref{fig_7_training_result}a. The inset depicts the ratio between the prediction and the label. From the trained ANN, the pulse-shapes are classified  according to the restoration coefficient as presented in Figure\,\ref{fig_7_training_result}b. The peak of the area-normalized pulse height decreases with the increasing restoration coefficient.

\begin{figure}[!h]
\includegraphics[width=15cm]{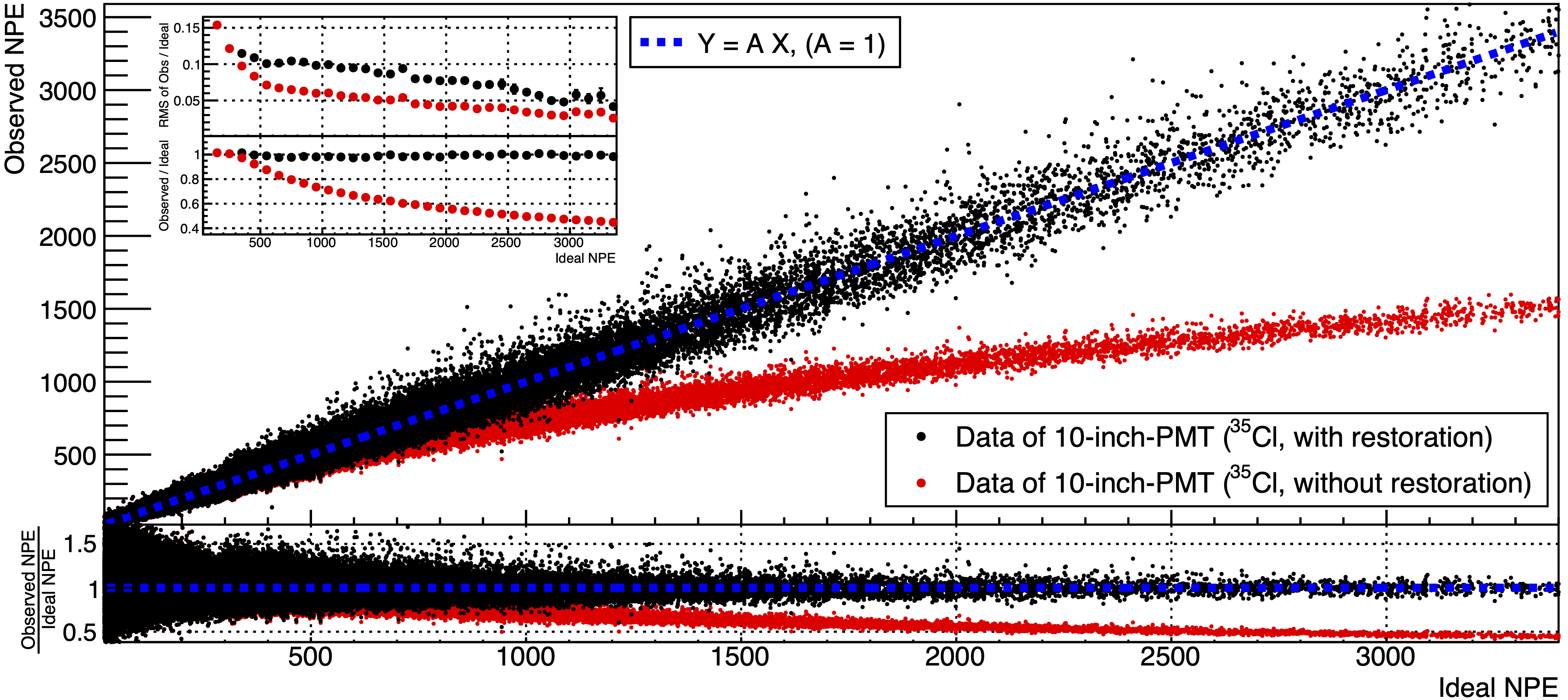}
\centering
\caption{Restoration of the saturation response of the PMT using the ANN model. The observed NPE of 10-inch-PMT without\,(with) restoration is shown in red\,(black). The blue dashed-line indicates the linear-model. The inset compares the bias and resolution of the PMT response with restoration and without restoration. The restoration was performed for NPE larger than 300.}
\label{fig_8_restoration}
\end{figure}

The ANN-predicted restoration coefficients obtained from the observed pulse-shapes were applied  to the test dataset. Figure\,\ref{fig_8_restoration} compares responses of the 10-inch-PMT with and without  restoration. A distinct saturation behavior appears as the NPE increases without restoration. The ANN-predicted restoration coefficient from the observed pulse-shape is multiplied with the NPE observed by the 10-inch-PMT to restore the decreased pulse-area. The restored response is compared with a linear response model$\rm\,(Y=AX)$ with the ideal slope$\rm\,(A=1)$. Figure\,\ref{fig_8_restoration} insets provide a comparison of the 10-inch-PMT response with\,(black points) and without\,(red points) restoration. Without restoration, a deviation from the ideal response appears in NPE larger than 300. The biased response was restored by the ANN from the prediction obtained by the pulse-shape classification. Due to the finite classification power, the root-mean-square\,(RMS) of the ideal/observed NPE ratio increased after the restoration. The RMS differences between the red and black decreased as the NPE increases.

\vspace{0.2cm}

\section{Summary \label{sec6}}

Neutrino events are typically reconstructed from the NPE observed by the PMTs mounted in the detector. Understanding the saturation behavior or a solid diagnosis of the linearity range of the PMT provides useful information for the correct reconstruction of the events. Previous studies were primarily focused on the PMT saturation behavior based on the absolute pulse-height or absolute-pulse-area.

In this study, the saturation behavior of a PMT was investigated by focusing on the correlation between two saturation responses, distortion of the pulse-shape, and decrease in the pulse-area. Comparing the observed pulse-shapes with the observed NPE provided useful information for an \textit{in-situ} diagnosis of the pulse-area decrease. The observed correlation between the two saturation responses was employed to train the ANN. The trained ANN predicted the decreased ratio of the pulse-area from the observed pulse-shape. The predicted restoration coefficient was applied to the observed NPE to restore the ideal case irrelevant to the pulse-area decrease. The restored linearity facilitates the correct reconstruction of the event in the saturation region for PMT-based detectors.

\section*{Acknowledgments}
This work was  supported by grants from the National Research Foundation\,(NRF) of the Korean Government\,(2022R1A2C1006069, 2022R1A5A1030700, 2022R1I1A1A01064311, 2018R1D1A1B07045812). We are very grateful for the support provided by the Center for Precision Neutrino Research at the Chonnam National University.

\vspace{0.2cm}
\noindent

\let\doi\relax

\newpage
\appendix
\section*{Appendix}
\section{Enlarged Insets for Appendix}
\label{sec:Appendix}

\begin{figure}[h]
\includegraphics[width=4.95cm]{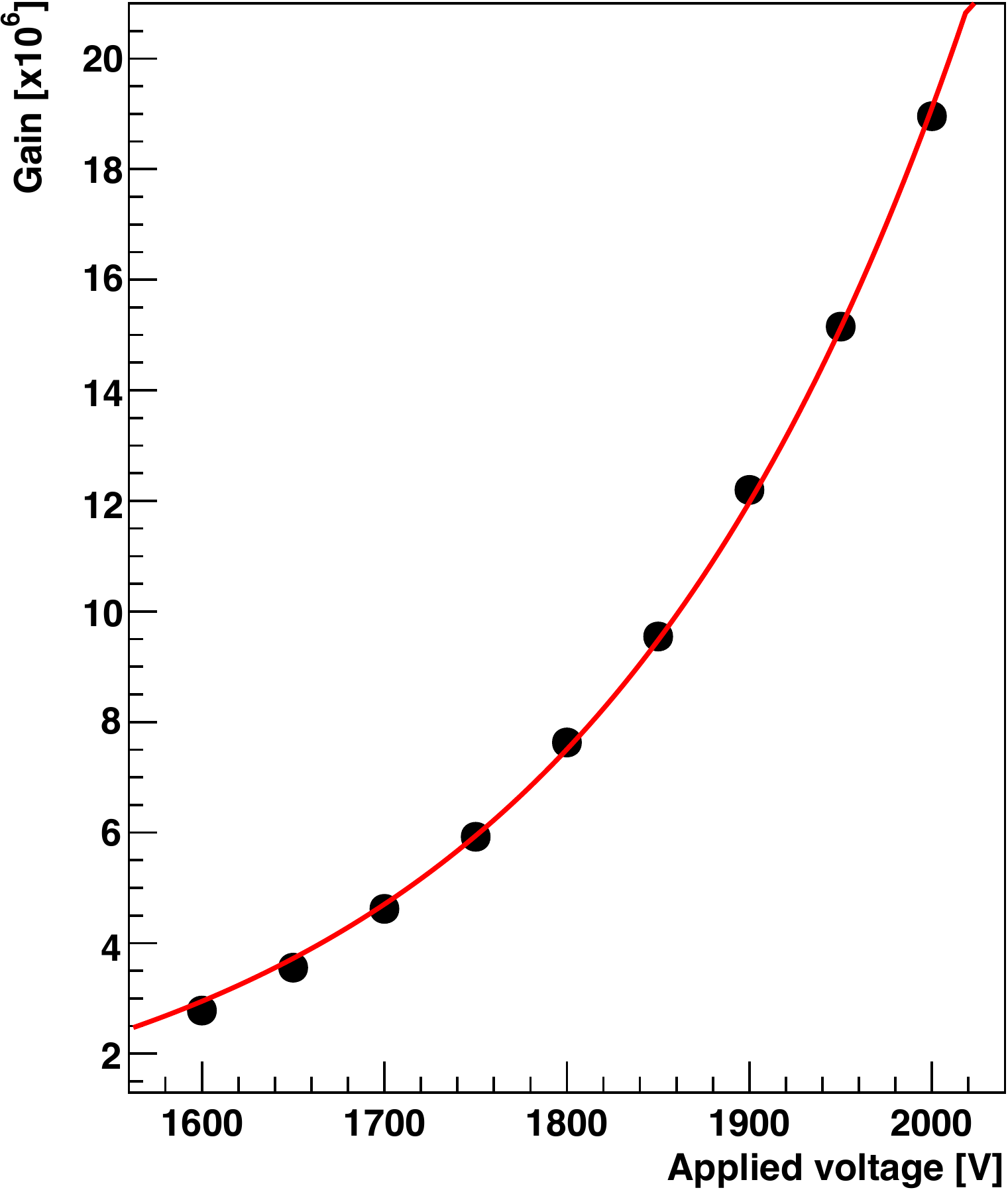}\,
\includegraphics[width=4.95cm]{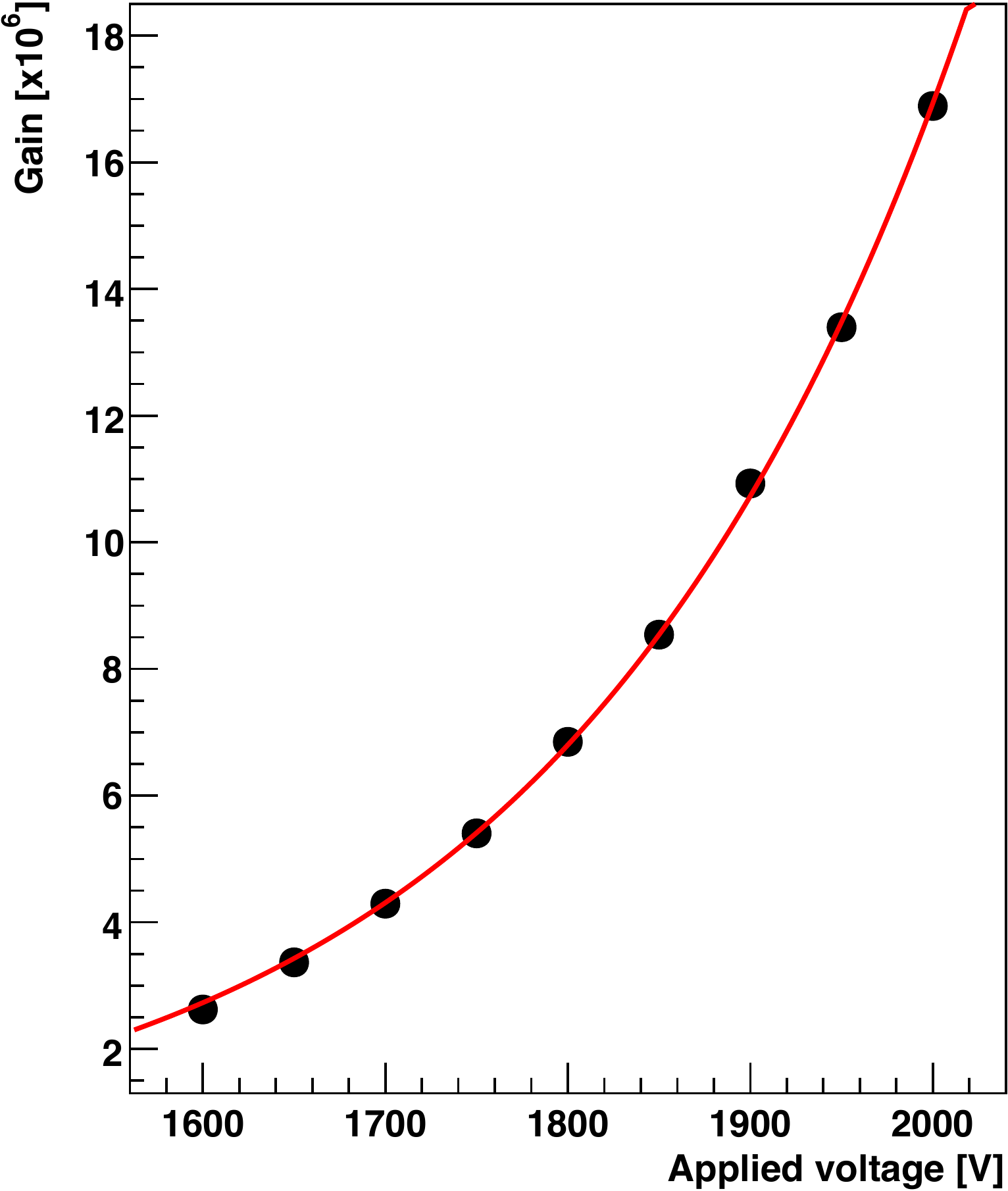}\,
\includegraphics[width=4.95cm]{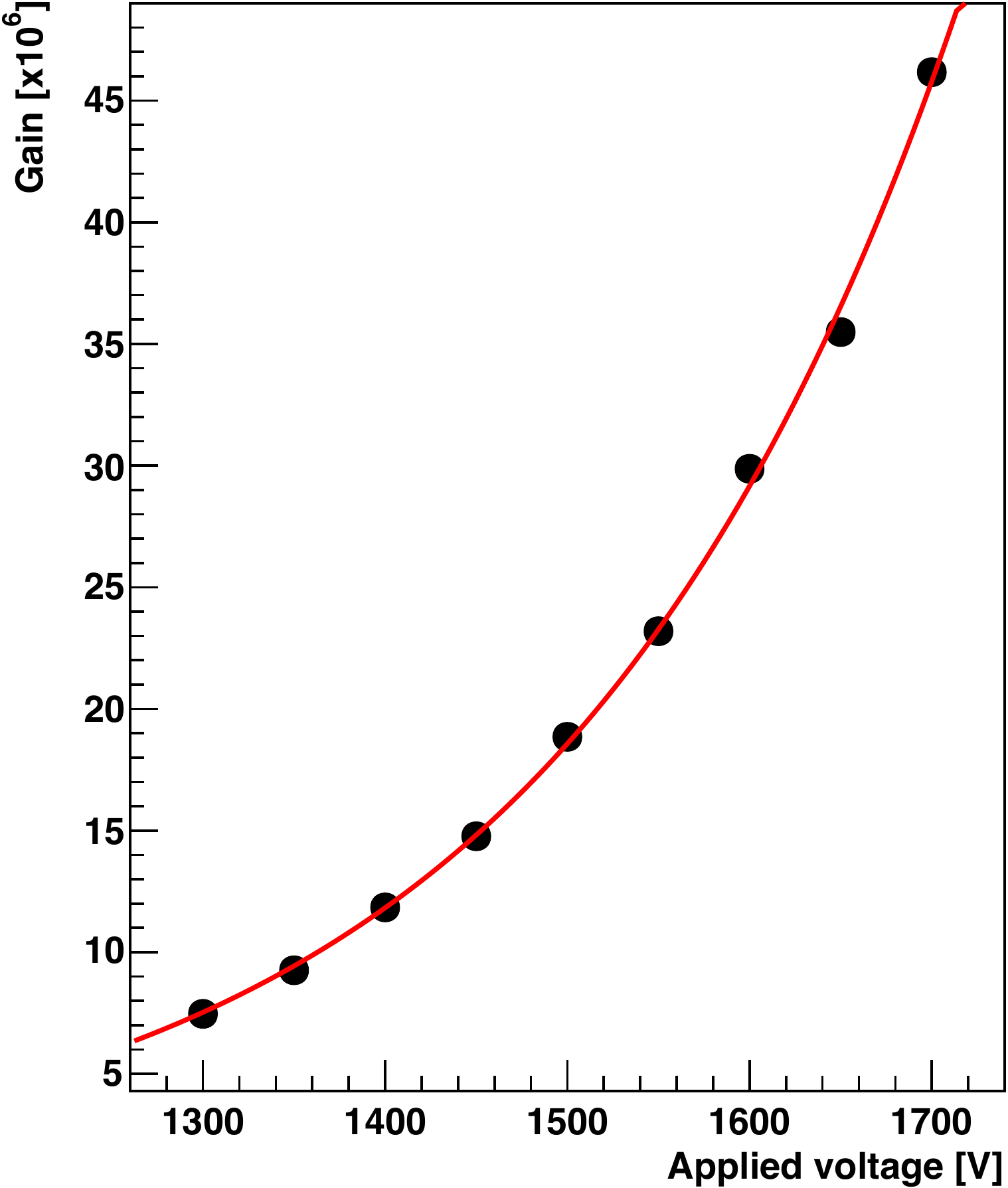}
\centering
\caption{Enlarged insets from Figure\,\ref{fig_3_gain}. The exponential increase in the observed gain with the increasing applied voltage for each PMT is presented. The gain curves for the 2-inch-A, 2-inch-B and 10-inch-PMT are displayed on the left, middle and right, respectively.} 
\label{fig_3_gain_A}
\end{figure}

\begin{figure}[h]
\centering
\includegraphics[width=7.5cm]{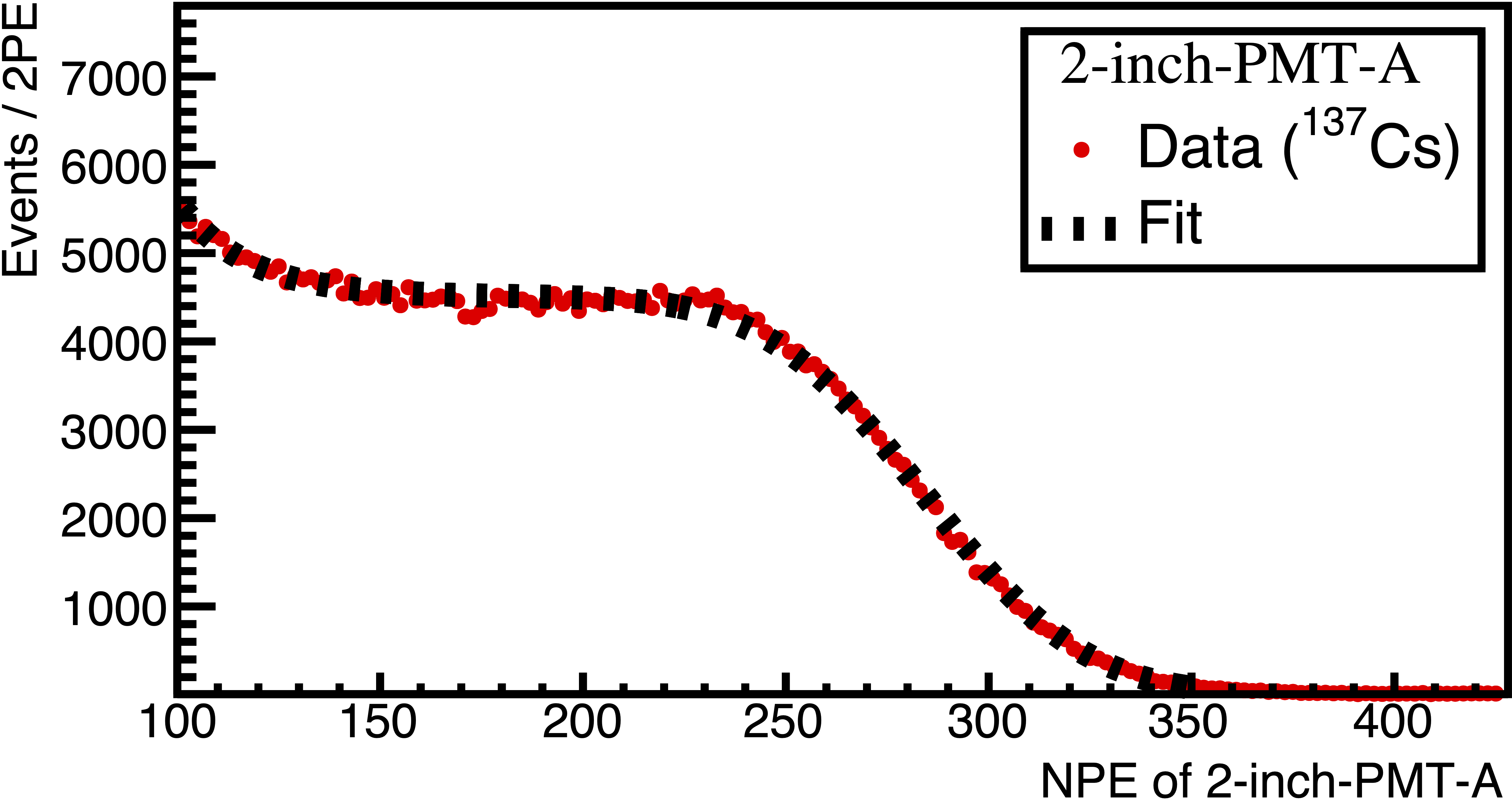}\,
\includegraphics[width=7.5cm]{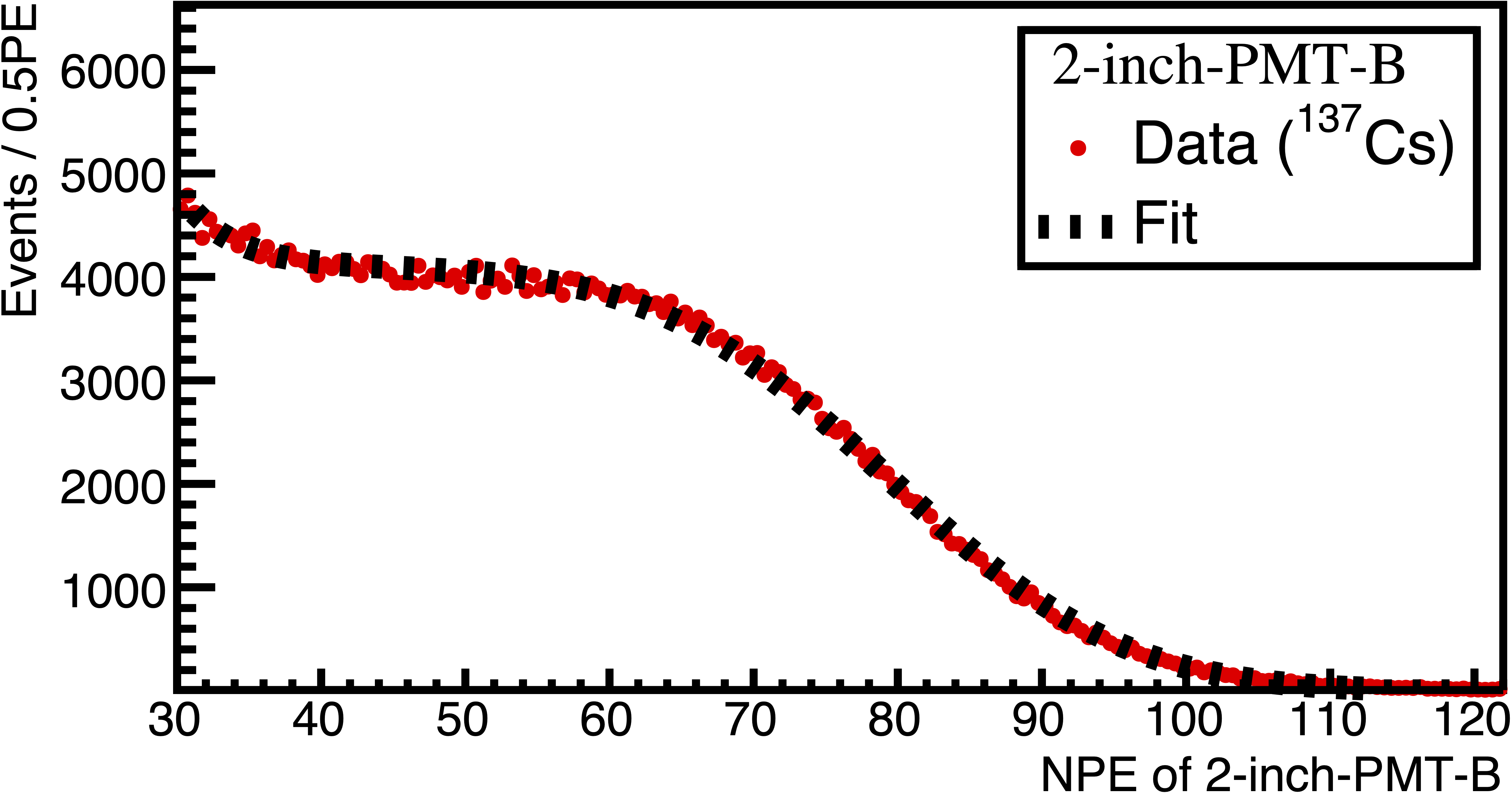}\,\newline
\includegraphics[width=9cm]{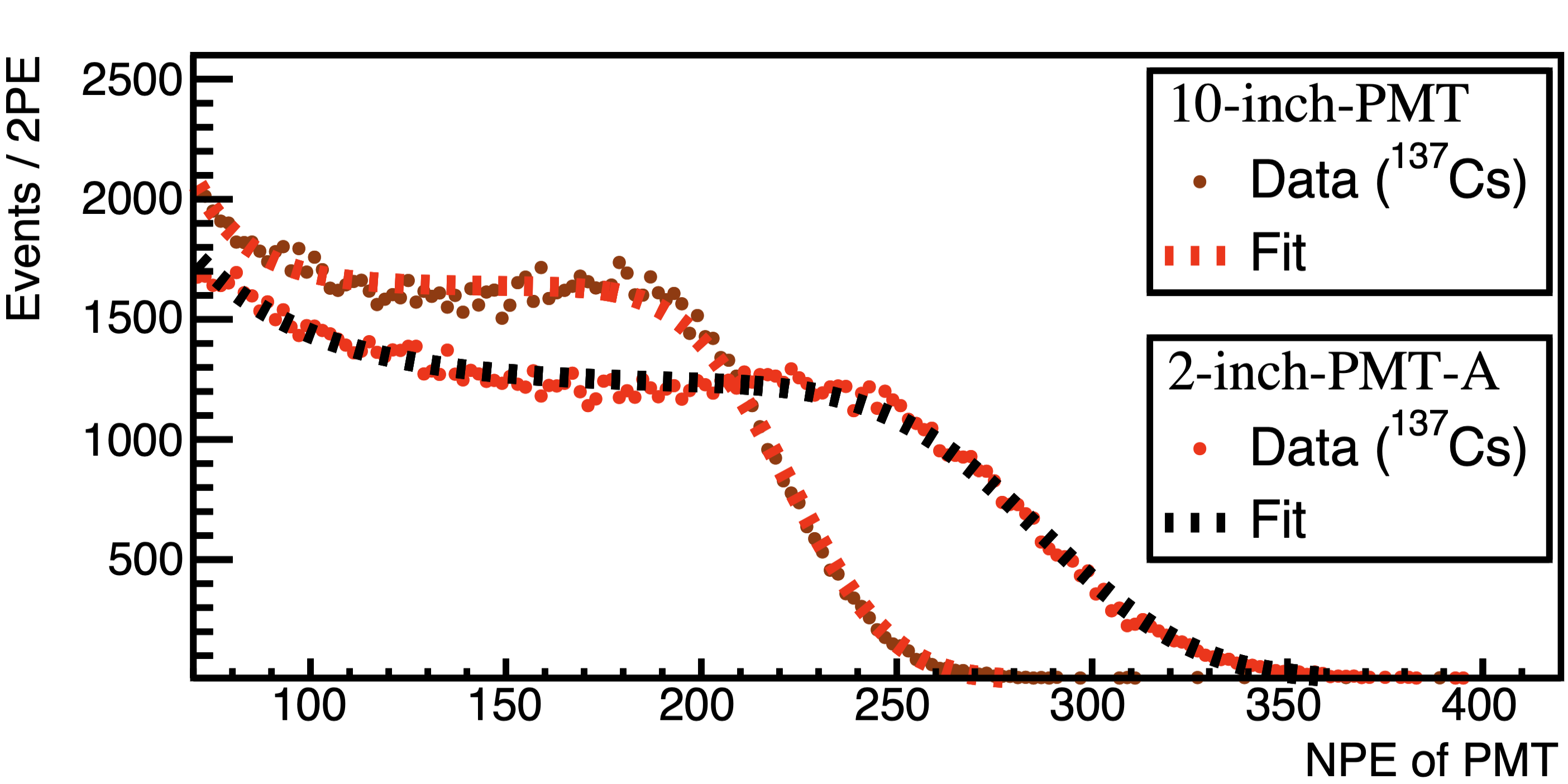}\;
\centering
\caption{Enlarged insets from Figure\,\ref{fig_4_PE_response}. One-dimensional projections of the red data points from Figure\,\ref{fig_4_PE_response}(a)\,(upper) and Figure\,\ref{fig_4_PE_response}(b)\,(lower), obtained using a $^{137}{\rm Cs}\,(0.66 \rm\, MeV)$  $\gamma$-ray source. The NPE distributions observed by each PMT are fitted with an empirical Error\,+\,Exponential function to determine the Compton edges.}
\label{fig_4_PE_response_A}
\end{figure}

\begin{figure}[h]
\includegraphics[width=10cm]{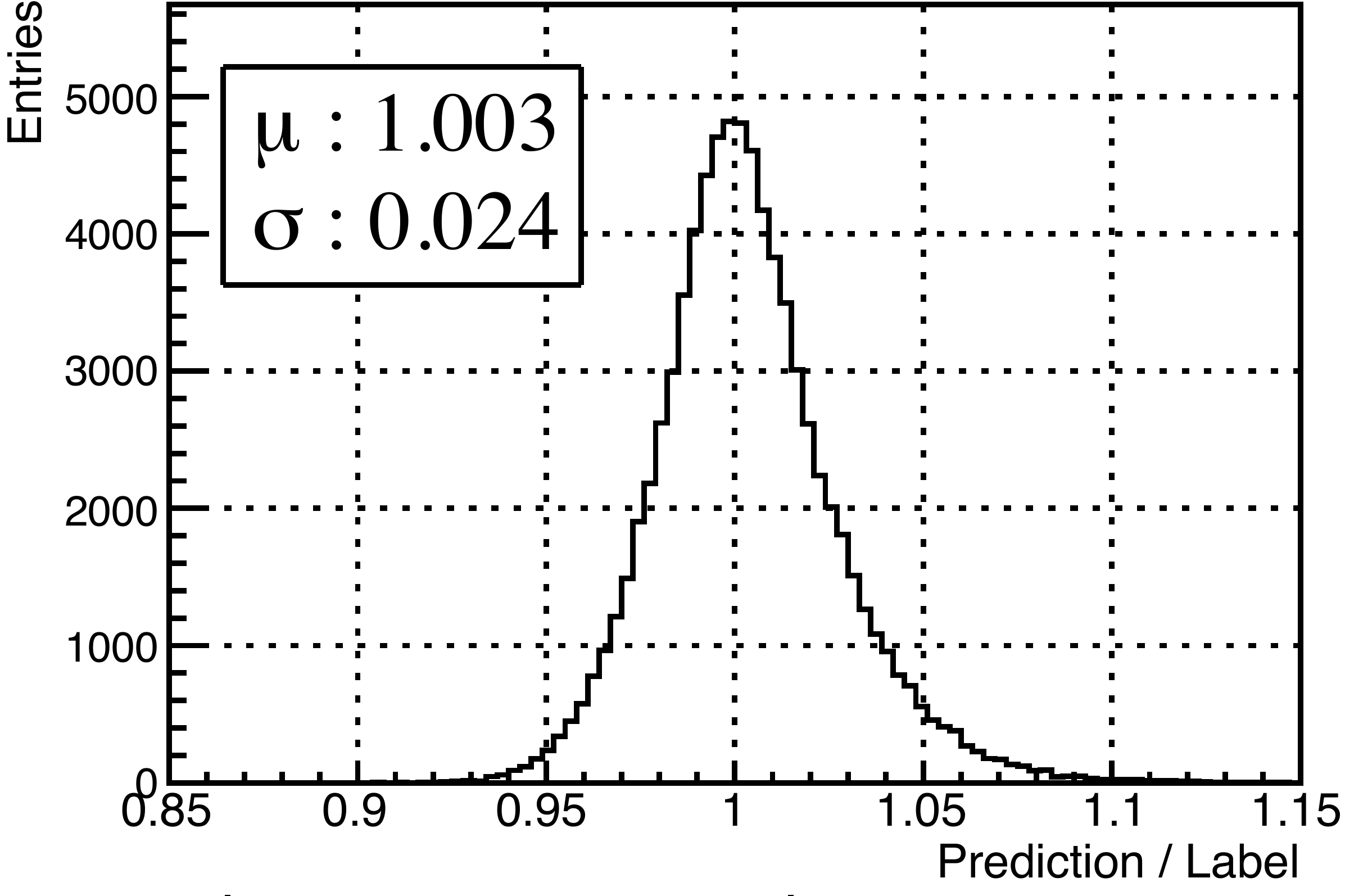}
\centering
\caption{Enlarged inset from Figure\,\ref{fig_7_training_result}(a). One-dimensional histogram of the ratio between the ANN prediction and the label of the restoration coefficients for the decreased pulse-area at the test dataset.}
\label{fig_7_training_result_A}
\end{figure}

\vspace{-1cm}

\begin{figure}[h]
\includegraphics[width=13cm]{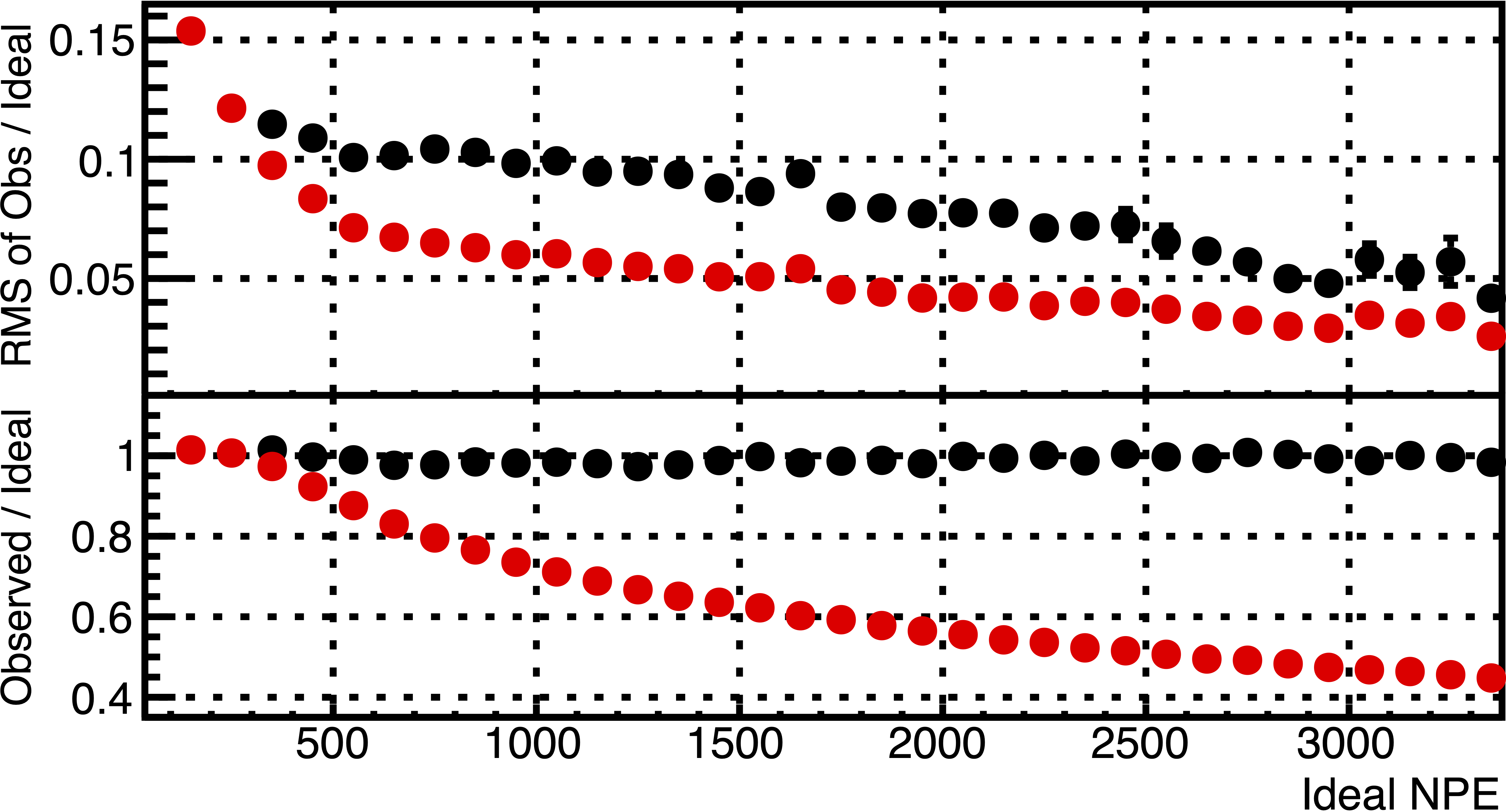}\,
\centering
\caption{Enlarged inset from Figure\,\ref{fig_8_restoration}. Comparison of the resolution and bias response of the 10-inch-PMT with restoration\,(black points) and without restoration\,(red points).  The resolution and bias response are represented by the Observed/Ideal ratio and the RMS of the Observed/Ideal ratio, respectively. }
\label{fig_8_restoration_A}
\end{figure}

\end{document}